\documentclass[a4paper]{llncs}

\usepackage{amsmath,amssymb}

\usepackage{xspace}

\usepackage[inline]{enumitem}

\usepackage{tikz}

\usepackage{cite}

\usepackage{url}

\usepackage{fullpage}

\newcommand{\rce}{\textsc{RCE}\xspace}
\newcommand{\python}{\text{Python}\xspace}
\newcommand{\java}{\text{Java}\xspace}
\newcommand{\fortran}{\text{Fortran}\xspace}
\newcommand{\matlab}{\text{Matlab}\xspace}

\newcommand{\cpp}{\text{C++}\xspace}

\newcommand{\digitalx}{\text{Digital-X}\xspace}

\newcommand{\freacs}{\text{FrEACs}\xspace}
\newcommand{\triad}{\text{TRIAD}\xspace}
\newcommand{\pegasus}{\text{PEGASUS}\xspace}
\newcommand{\clava}{\text{CLAVA}\xspace}
\newcommand{\inteeverII}{\text{INTEEVER II}\xspace}
\newcommand{\trak}{\text{TraK}\xspace}
\newcommand{\veuII}{\text{VEU II}\xspace}\newcommand{\holiship}{\text{HOLISHIP}\xspace}

\newcommand{\dlr}{\text{DLR}\xspace}

\newcommand{\MS}{\ensuremath{\mathrm{MS}}\xspace}
\newcommand{\AD}{\ensuremath{\mathrm{AD}}\xspace}
\newcommand{\SE}{\ensuremath{\mathrm{SE}}\xspace}
\newcommand{\F}{\ensuremath{\mathrm{F}}\xspace}
\newcommand{\R}{\ensuremath{\mathrm{R}}\xspace}

\newcommand{\ssh}{\text{SSH}\xspace}

\usetikzlibrary{calc,shapes,fit,decorations.pathmorphing,decorations.pathreplacing,patterns,arrows}
\tikzset{every edge/.style = {-stealth, draw}}
\tikzset{thick,>=stealth, shorten >=1pt}

\colorlet{darkgray}{black!15}
\colorlet{lightgray}{gray!5}

\pgfdeclarelayer{background}
\pgfdeclarelayer{foreground}
\pgfsetlayers{background,main,foreground}

\begin{document}

\title{RCE: An Integration Environment for Engineering and Science}

\author{%
	Brigitte Boden \and%
	Jan Flink \and%
	Niklas F{\"o}rst \and%
	Robert Mischke \and%
	Kathrin Schaffert \and \\%
	Alexander Weinert \and%
	Annika Wohlan \and%
	Andreas Schreiber%
}

\institute{
	German Aerospace Center (DLR), Simulation and Software Technology, \\
	Linder H{\"o}he, 51147 K{\"o}ln, Germany \\
	\email{firstname.lastname@dlr.de}
}

\maketitle

\begin{abstract}
Engineering of complex systems such as air- and spacecraft is a multidisciplinary effort that requires the collaboration of engineers from a multitude of specializations working in concert.
Typically, each engineer uses one or more specialized software tools to analyze some data set and passes, in an ad-hoc manner, the results on to their colleagues who require these results as input for their respective tools.
This process is time-consuming, error-prone, and not replicable.

To alleviate this problem, we present \rce (Remote Component Environment), an open-source application developed primarily at \dlr, that enables its users to intuitively integrate disciplinary tools, to define dependencies between them via an easy-to-use graphical interface, and to execute the resulting multidisciplinary engineering workflow.
All data produced are stored centrally for provenance, subsequent analysis, and post-processing.
Hence, \rce makes it easy for collaborating engineers to contribute their individual disciplinary tools to a multidisciplinary design or analysis, and simplifies the analysis of the workflow's results. 
\end{abstract}

\section{Motivation and Significance}
\label{sec:motivation_and_significance}

Designing and evaluating modern systems requires collaboration among experts from a wide range of disciplines.
Each of these experts uses highly specialized software tools to produce or investigate a design artifact such as the shape of a workpiece, its aerodynamic properties, or the cost for producing it.
Orchestrating the runs of these tools as well as collecting, distributing, and archiving all intermediate data imposes a significant organizational overhead on the design process.
\rce (Remote Component Environment) \cite{rce_zenodo} enables engineers and scientists to construct automated workflows consisting of numerous such software tools, to execute these workflows on a distributed network of compute nodes, and to collect all relevant artifacts for further analysis.

As an example, consider a team of engineers working to determine whether to construct a novel airplane wing out of steel, aluminum, or carbon.
There is also a project leader who coordinates the activities involved in determining the optimal material.
To this end, for each of the three materials, the engineers compute both the effectiveness of the resulting wing as well as its production cost.
For the sake of simplicity, we assume the effectiveness of a wing shape is determined solely based on the lift it produces.

Each of the above materials may induce different constraints on the possible shapes.
Hence, the project leader first obtains a description of these constraints from the collaborating material scientist.
She then passes these constraints on to a team of engineers that determine, for each material, an optimal feasible shape that produces maximal lift.
They do so by iteratively designing wing shapes and evaluating the lift produced by the respective shape, thus implementing a simple optimization loop.
Having determined an optimal wing shape for a given material, the project leader then forwards the optimal design to another department that estimates the production cost.
We illustrate the flow of information between the participants involved in the design of the wing in Figure~\ref{fig:initial-workflow}.

\begin{figure}
\centering
\begin{tikzpicture}[thick,xscale=.95]

\tikzset{
	artifact/.style = {
		align=center,
		rounded corners,
	},
	collaborator/.style = {
		align=left,
		anchor=west
	},
	data flow/.style = {
		draw,rounded corners=.1cm,-stealth
	}
}

\node[artifact] (1) at (1.2,-1) {\footnotesize Const.};
\node[artifact] (2) at (2.1,0) {\footnotesize Const.};
\node[artifact] (3) at (3,-2) {\footnotesize Shape};
\node[artifact] (4) at (3.8,-3) {\footnotesize Lift};
\node[artifact] (5) at (4.6,-2) {\footnotesize Shape};
\node[artifact] (6) at (5.4,-3) {\footnotesize Lift};
\node[align=center] (7) at (6.2,-2) {\footnotesize \dots};
\node[align=center] (8) at (6.8,-3) {\footnotesize \dots};
\node[artifact] (9) at (7.8,-2) {\footnotesize Shape};
\node[artifact] (10) at (8.8,0) {\footnotesize Shape};
\node[artifact] (11) at (9.8,-4) {\footnotesize Cost Est.};
\node[artifact] (12) at (11,0) {\footnotesize Cost Est.};

\foreach \y in {-.5,-1.5,-2.5,-3.5} {
	\draw[gray] (-1.85,\y) -- (12,\y);
}

\path[data flow] (1) -| (2);
\path[data flow] (2) -| (3);
\path[data flow] (3) -| (4);
\path[data flow] (4) -| (5);
\path[data flow] (5) -| (6);
\path[data flow] (8) -| (9);
\path[data flow] (9) -| (10);
\path[data flow] (10) -| (11);
\path[data flow] (11) -| (12);

\node[collaborator] (project-leader) at (-1.8,0) {\footnotesize Project \\[-.1cm] \footnotesize Leader};
\node[collaborator] at (-1.8,-1) {\footnotesize Materials \\[-.1cm] \footnotesize Science};
\node[collaborator] (structural-engineering) at (-1.8,-2) {\footnotesize Structural \\[-.1cm] \footnotesize Engineering};
\node[collaborator] at (-1.8,-3) {\footnotesize Aerody-\\[-.1cm] \footnotesize namics};
\node[collaborator] (finance) at (-1.8,-4) {\footnotesize Finance};

\coordinate (separator-x) at (structural-engineering.east);
\coordinate (separator-y-north) at (project-leader.north);
\coordinate (separator-y-south) at (0,-4.5);

\path[draw,gray] (separator-x |- separator-y-north) -- (separator-x |- separator-y-south);

\begin{pgfonlayer}{background}

\draw[draw=none,fill=darkgray] (-1.85,.5) rectangle (separator-x |- 0,-.5);
\fill[draw=none,left color=darkgray, right color = white]
	(separator-x |- 0,.5) rectangle (12,-.5);
	
\draw[draw=none,fill=lightgray] (-1.85,-.5) rectangle (separator-x |- 0,-1.5);
\fill[left color=lightgray, right color = white]
	(separator-x |- 0,-.5) rectangle (12,-1.5);
	
\draw[draw=none,fill=darkgray] (-1.85,-1.5) rectangle (separator-x |- 0,-2.5);
\fill[left color=darkgray, right color = white]
	(separator-x |- 0,-1.5) rectangle (12,-2.5);
	
\draw[draw=none,fill=lightgray] (-1.85,-2.5) rectangle (separator-x |- 0,-3.5);
\fill[left color=lightgray, right color = white]
	(separator-x |- 0,-2.5) rectangle (12,-3.5);
	
\draw[draw=none,fill=darkgray] (-1.85,-3.5) rectangle (separator-x |- 0,-4.5);
\fill[draw=none,left color=darkgray, right color = white]
	(separator-x |- 0,-3.5) rectangle (12,-4.5);
	
\end{pgfonlayer}

\end{tikzpicture}
\caption{Process for designing and evaluating a wing shape.}
\label{fig:initial-workflow}
\end{figure}

Each step of such a design process uses one or multiple specialized software tools.
Without a common software to coordinate data management, input and output files are typically transmitted between these tools using ad hoc methods like email, flash drives, or shared network drives.
Even within a single design process, both the programming languages used for implementing these tools as well as the environments required for executing them are often heterogeneous.
In our example, engineers may compute the shape constraints using \python~\cite{python} running on a desktop computer, whereas the wing shape may be constructed using \matlab~\cite{matlab} on a compute cluster.
Finally, the aerodynamic evaluation may be performed using a simulation implemented in \fortran~\cite{fortran} using a GPU cluster while the estimate of production costs may be implemented in \java~\cite{java}.

There are several obstacles in executing such a multidisciplinary design process.
Not only do the participants from different disciplines have to agree on methods for collaborating, possibly across the boundaries of business units, but also, on a more technical level, on data formats for data exchange.
Moreover, they have to manually invoke their respective tools as well as subsequently collect, disseminate, and archive the resulting data.
Even if this process is automated at all, this automation usually consists of custom solutions that are not easily reusable for other processes.

\rce is an open-source software that enables scientists to overcome these recurrent obstacles when executing processes involving multidisciplinary software tools running in diverse runtime environments.
While \rce cannot solve the social and disciplinary challenges outlined above, it supports its users by resolving the technical issues regarding this process.
For the remainder of this work, we refer to the technical component of a multidisciplinary design process as a multidisciplinary workflow.
In order to execute a workflow, \rce lets the user grant other users in the same network access to selected tools on their local machine, allowing the remote users to execute the tool on demand, while the tool owner retains full control over the implementation files and runtime environment of the tool.
\rce furthermore enables the users to construct and execute automated workflows comprised of both local tools as well as those published by other users in the same network.
For the construction of such workflows, \rce supplies the user with an intuitive graphical interface.
During automated workflow execution, \rce requires no further user input and orchestrates both the distribution of inputs and outputs between the involved tools and their invocation on the respective machines.
Finally, it allows users to monitor the artifacts and console output accrued during the execution of the individual tools.
All execution data is also automatically persisted, allowing access to and analysis of this data both during and after a workflow's execution.

In doing so, \rce facilitates cooperation between engineers and researchers of widely differing fields, enables them to leverage their domain expertise to simulate and analyze complex systems.
\rce furthermore supports the engineers and researchers in discovering and tracing relationships between the input parameters of the workflow and its results.
It has been used in many projects at \dlr and users have consistently reported that in addition to enabling larger and more complex workflows, \rce has also greatly increased their efficiency.

\paragraph*{Related Work}
\label{sec:introduction:relatedwork}
While, to the best of our knowledge, no other software shares the concept and aim of \rce, there exists a number of tools that have similarities with \rce.
Here we highlight some of the most prominent of these tools and their main differences.

One software whose aims most align with \rce is ModelCenter by Phoenix Integration~\cite{ModelCenter}. 
While ModelCenter is proprietary, \rce is open-source and available free of charge.

OpenMDAO~\cite{GrayHwanMartinsMooreNaylor2019} is an open-source framework for multidisciplinary optimization.
While earlier versions of OpenMDAO had some features in common with \rce, starting with version~2 the developers focused solely on optimization. 
Hence, current versions of \rce and OpenMDAO have orthogonal feature sets.

Finally there exist Apache Nifi~\cite{ApacheNifi} and Knime~\cite{Knime}. 
While both feature intuitive graphical editors, the main focus of these tools lies on the domain of data science. 
\rce, in contrast, focuses on the field of multidisciplinary engineering.

\paragraph*{Paper Structure}
\label{sec:introduction:paperstructure}
In Section~\ref{sec:software_description} we describe the functionalities of \rce from the users' point of view as well as the underlying architecture.
In Section~\ref{sec:illustrative_examples} we present a comprehensive example of the usage of \rce in a fictitious aviation project.
In Section~\ref{sec:impact} we give several examples of the usage of \rce in research projects at \dlr before summarizing our results in Section~\ref{sec:conclusion} and giving an outlook at future work.

\section{Software Description}
\label{sec:software_description}

We first describe the use cases supported by \rce in Section~\ref{sec:software_description:functionalities}, before we discuss the implementation of the features enabling these use cases in Section~\ref{sec:software_description:architecture}.

\subsection{Software Functionalities}
\label{sec:software_description:functionalities}

\rce is a general tool for engineers and scientists from a wide range of disciplines to create and execute distributed workflows as well as evaluate the obtained results.
Hence, \rce supports a wide range of use cases.

As a first use case, consider a single engineer who has to orchestrate the execution of multiple tools in her daily work and wants to automate this process.
To this end, she can integrate external tools for calculations, simulations, and evaluations using \rce.
\rce provides a graphical wizard that guides the user through the tool integration process.

During integration, the user defines inputs and outputs of the tool using the native datatypes of \rce, which comprise typical primitive data types such as Booleans, integers, and floating point numbers as well as more general ones such as files and directories.
The user can also define script sections that are executed before and after each tool run (called pre- and post-scripts), which is typically used for the conversion of input transmitted by \rce into tool-specific layouts and, vice versa, for the conversion of tool-specific output data into more common formats before they are handed over to \rce.
Finally, the user defines the commands to execute the underlying tool given the previously defined inputs.

\begin{figure}
	\centering
	\input{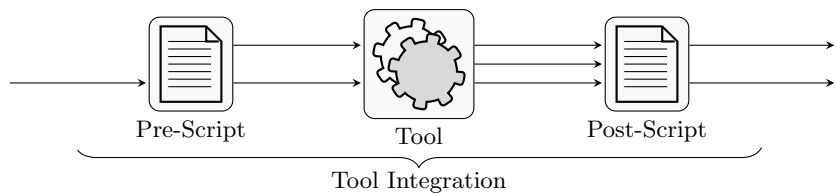}
	\caption{A conceptual view of the integration of a tool into \rce.}
	\label{fig:integration}
\end{figure}

We illustrate the integration of a tool into \rce in Figure~\ref{fig:integration}.
In this example, the pre-script takes one parameter that is supplied by \rce and provides a second, hard-coded parameter.
The integrated tool is then called with these two parameters and provides three output values.
Out of these output values, the post-script discards one and passes the other two back to \rce.

\begin{figure}
	\centering
	\includegraphics[width=\textwidth]{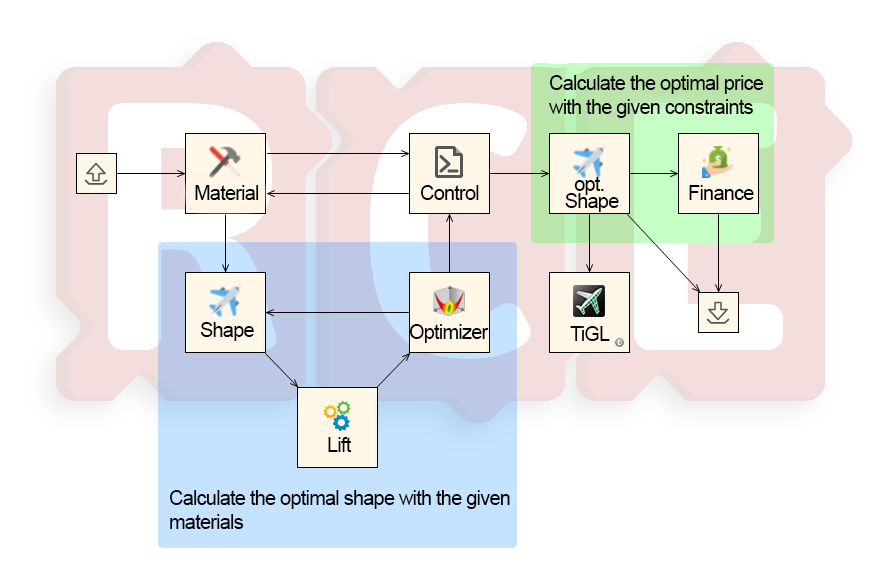}
	\caption{A workflow in \rce. The fontsize of the labels has been increased for the sake of readability.}
	\label{fig:rce-workflow}
\end{figure}

In addition to the user-integrated tools, \rce provides a number of predefined tools which supply basic and often-needed functionalities such as controlling the flow of data through the workflow, reading and extracting data from XML files, executing user-supplied Python scripts, as well as a number of tools that provide mathematical and statistical methods to evaluate the incoming data.
Furthermore, \rce supplies components for the construction of design-, evaluation- and optimization loops, most prominent among them an integration of the well-known and versatile Dakota framework~\cite{dakota} for optimization.
The user can mix and match both pre-integrated tools and user-integrated ones to construct complex workflows in the graphical workflow editor of \rce.
As an example, a user may construct a workflow for determining the optimal material for the construction of an airplane wing as described in Section~\ref{sec:motivation_and_significance}.
We show such a workflow in Figure~\ref{fig:rce-workflow}.

After constructing the workflow, the user can start its execution.
As a first step, \rce triggers the execution of all tools without inputs.
Once a tool has terminated, \rce collects its outputs as defined by its integration and computed by its post-script, and passes them on to the tools whose inputs are connected to the outputs of the terminated tool.
\rce then executes the succeeding tools. Once it has collected all inputs defined by the integration of a tool, it executes the respective tool and again collects its outputs to pass them to subsequent tools.
This is continued until no tool has produced any further output data, which is typically the case either because a linear workflow has executed all of its tools, or because a main evaluation loop has reached a certain end criterion (e.g., convergence).

Execution via \rce is also possible for tools that are under active development.
Instead of ``freezing'' a certain version of software tools and executing this version every time the workflow is executed, \rce instead uses the version currently installed on the machine.
Hence, there is no additional overhead in deploying new versions of actively developed tools to \rce.
Instead, \rce executes the tool version that has been most recently deployed.
Once such a tool has stabilized, the user may provide both a ``development'' version and a ``stable'' version in parallel.
Thus, one set of users may use the stable version of a tool, while others gain immediate access to one or more development versions at the same time.

Now consider the use case that the user wants to execute some parts of the workflow on her local machine, while she wants to execute others on a compute cluster or some remote cluster equipped with GPUs.
To this end, she can connect her local instance of \rce with one or multiple instances running on remote machines reachable via a local network or the Internet.
She may configure the remote instances either via a graphical interface, if the remote machine offers one, or via the built-in command-line interface of \rce, which is also reachable directly via an \ssh connection.

The engineer may also connect to instances of \rce controlled by other engineers via the same mechanism.
Users can publish their tool integrations via the network that was created this way, allowing remote users to use the locally integrated tools in their workflows.
When constructing a workflow, remote and local components appear side-by-side in the graphical editor and are indistinguishable from one another.
This allows the user to construct the logic of the workflow without having to determine its execution configuration at the same time.

For execution of the resulting workflow comprising both local and remote components, the latter ones are not copied to the machine executing the workflow, but they are instead executed on the machine on which they are integrated.
\rce supplies them with the required input data via the network connection, retrieves their output after termination and transmits that data to the machine coordinating the workflow's execution.

In order to simplify the setup of \rce networks for collaboration, \rce instances can also be configured to run as a so-called relay.
Other \rce instances can then connect to this instance, which acts as an intermediate, forwarding relevant information between all instances as needed.
As \rce's network structure is decentralized, any number of relays can be combined, which is often used to first connect the \rce instances within a working group, and to then assemble these groups into larger networks.

If the engineer using \rce has integrated all tools involved in her workflow onto remote instances, she may want to shut down her local machine during the execution of the workflow.
To this end, upon execution of a workflow, the user may designate some instance to be the workflow controller, which orchestrates the distribution of data among the involved instances of \rce.
This is particularly useful for the execution of long-running workflows, as the user may start the workflow on their local machine and designate a server as the workflow controller, allowing them to shut down their local machine after the start of the workflow.

Finally, consider the case that the engineer wants to inspect the output of individual executions of some particular tool that has been invoked during the execution of the workflow.
To this end, all data, i.e., all inputs to and outputs from all components accrued during a workflow run are collected by the workflow controller and can be monitored both during execution of the workflow and after its completion by all \rce instances connected to the workflow controller.
This provides a huge benefit for the collaborative analysis of the results of the workflow over the existing ad-hoc dissemination of such results.

In practice, the roles of the \rce instances involved in the execution of a workflow may not be as clearly separated as in the use cases discussed above.
A single \rce instance can be configured to display a user interface for the construction of workflows, to publish integrated tools and act as a workflow controller, and to merge connected networks, or to do any combination of these tasks.

\subsection{Software Architecture}
\label{sec:software_description:architecture}

The user base of \rce is mainly comprised of engineers and scientists who are not primarily interested in designing workflows, but rather in performing multidisciplinary analyses.
Hence, we provide users with a software package that is \emph{(i) easy to deploy and configure} and \emph{(ii) features very few external dependencies}.
Moreover, such analyses are usually performed on specialized hardware that is shared with other users, e.g., a compute cluster.
Hence, the users of \rce may not have complete administrative control over the machine they use.
Thus, another main driver behind the architectural decisions regarding \rce is the requirement that the resulting software must be able to \emph{(iii) run on a wide range of machines with (iv) minimal privileges}.
As \rce is not developed for use in a single domain, but rather as a general tool for the development and execution of workflows, we require a \emph{(v) modular architecture} that is amenable to the development of additional modules as required by certain projects.
Finally, since users must be able to configure networks of \rce instances, we require an architecture that allows for \emph{(vi) easy communication} between connected instances.

To achieve platform independence, we opted to implement \rce in \java~\cite{java}.
As we furthermore rely on the platform-independent Eclipse Rich Client Platform (RCP)~\cite{eclipse-rcp}, we can provide executable artifacts for Windows and Linux while only maintaining a single code base.
By also distributing \rce as a simple zip file (besides more specialized installation packages), which contains all dependencies of \rce save for the actual Java runtime, we achieve the first, second, and third requirement listed above.
As \rce only requires a normal user account without elevated privileges, we furthermore achieve the fourth requirement above.

To achieve the fifth requirement above, \rce has a highly modular structure on the source code level.
Every significant functionality is accessible from other functional units via an explicit service interface.
This approach ensures a clear separation of code segments, improves maintainability, and facilitates testing.
We define these services via the OSGi Declarative Services standard~\cite{osgi-ds}, and manage them at runtime using the Eclipse Equinox OSGi implementation~\cite{equinox} included in RCP.

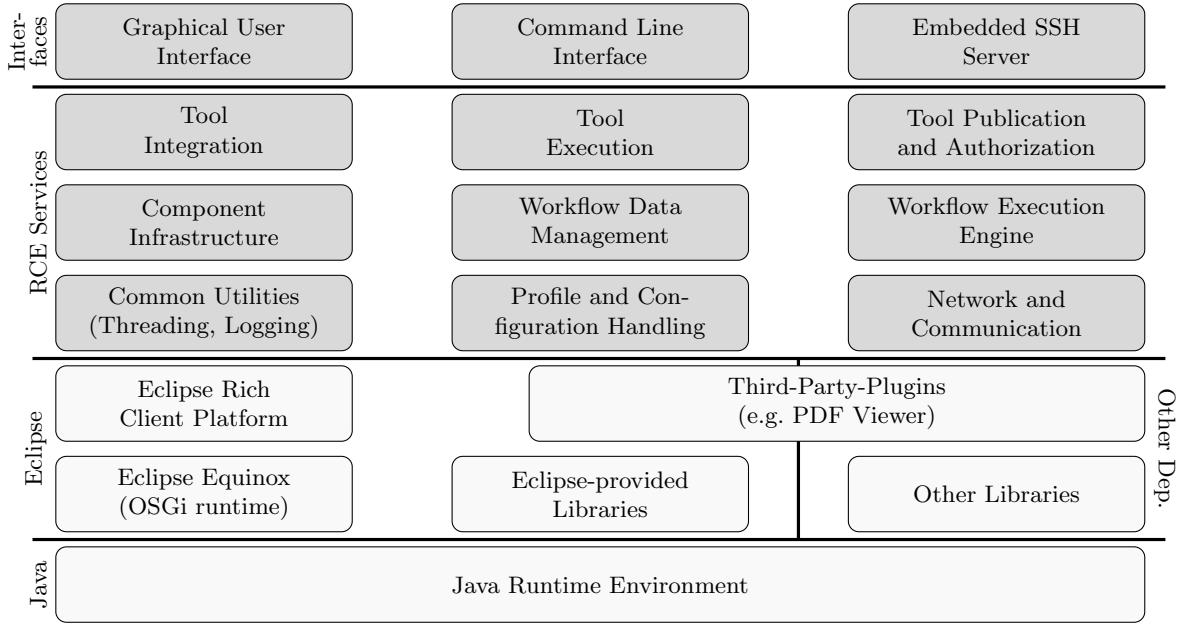
\begin{figure}
\centering
\begin{tikzpicture}[yscale=.6]

\definecolor{servicecolor}{HTML}{83bfe2}
\definecolor{dependencycolor}{HTML}{fde693}

\tikzset{
	servicebox/.style = {
		align=center, rounded corners,
		minimum width = 3.9cm,
		minimum height = 1cm,
		font=\footnotesize
	},
	dependency/.style = {
		servicebox,
		draw=black,fill=gray!5
	},
	service/.style = {
		servicebox,
		draw=black, fill=black!15
	},
	separator/.style = {
		black, very thick
	}
}

\node[dependency,minimum width=.9\textwidth] (jre)
	at (0,0)
	{Java Runtime Environment};
	
\node[dependency,anchor=east] (custom-lib)
	at (jre.east |- 0,2)
	{Other Libraries};
\node[dependency,anchor=west] (equinox)
	at (jre.west |- 0,2)
	{Eclipse Equinox\\(OSGi runtime)};
\node[dependency] (eclipse-lib)
	at (jre |- 0,2)
	{Eclipse-provided \\ Libraries};
\node[dependency,anchor=west] (rcp)
	at (equinox.west |- 0,4)
	{Eclipse Rich \\ Client Platform};
\node[dependency,anchor=east,minimum width=8.1cm] (third-party-plugins)
	at (custom-lib.east |- 0,4)
	{Third-Party-Plugins\\(e.g. PDF Viewer)};
	
\node [service,anchor=west] (utilities)
	at (rcp.west |- 0,6)
	{Common Utilities \\ (Threading, Logging)};
\node [service] (profile-handling)
	at (eclipse-lib |- 0,6)
	{Profile and Con-\\ figuration Handling};
\node [service,anchor=east] (network)
	at (custom-lib.east |- 0,6)
	{Network and \\ Communication};
	
\node [service,anchor=west] (wf-infra)
	at (rcp.west |- 0,8)
	{Component \\ Infrastructure};
\node [service] (wf-dat-man)
	at (eclipse-lib |- 0,8)
	{Workflow Data \\ Management};
\node [service,anchor=east] (wf-exec-engine)
	at (custom-lib.east |- 0,8)
	{Workflow Execution \\ Engine};
	
\node [service,anchor=west] (tool-integration)
	at (rcp.west |- 0,10)
	{Tool \\ Integration};
\node [service] (tool-exec)
	at (eclipse-lib |- 0,10)
	{Tool \\ Execution};
\node [service,anchor=east] (tool-publication)
	at (custom-lib.east |- 0,10)
	{Tool Publication \\ and Authorization};
	
\node [service,anchor=west] (gui)
	at (rcp.west |- 0,12)
	{Graphical User \\ Interface};
\node [service] (cmd)
	at (eclipse-lib |- 0,12)
	{Command Line \\ Interface};
\node [service,anchor=east] (ssh)
	at (custom-lib.east |- 0,12)
	{Embedded SSH \\ Server};

\begin{pgfonlayer}{background}	
\draw[separator] ($(jre.north west) + (-.3cm, .15)$) -- ($(jre.north east) + (.3cm, .15)$);		
\draw[separator] ($(rcp.north west) + (-.3cm, .15)$) -- ($(third-party-plugins.north east) + (.3cm, .15)$);
\draw[separator] ($(tool-integration.north west) + (-.3cm, .15)$) -- ($(tool-publication.north east) + (.3cm, .15)$);
\coordinate (separator-x) at ($(eclipse-lib) ! .5 ! (custom-lib)$);
\coordinate (separator-y-low) at ($(jre.north west) + (-.3cm, .15)$);
\coordinate (separator-y-high) at ($(rcp.north west) + (-.3cm, .2)$);
\draw[separator] (separator-x |- separator-y-low) -- (separator-x |- separator-y-high);
\end{pgfonlayer}

\node[anchor=south,rotate=90] at (jre.west) {\footnotesize Java};
\node[anchor=south,rotate=90] at ($(equinox.west) ! .5 ! (rcp.west)$) {\footnotesize Eclipse};
\node[anchor=south,rotate=-90] at ($(third-party-plugins.east) ! .5 ! (custom-lib.east)$) {\footnotesize Other Dep.};
\node[anchor=south,rotate=90] at (wf-infra.west) {\footnotesize RCE Services};
\node[anchor=south,rotate=90,align=center] at (gui.west) {\footnotesize Inter- \\[-.1cm] \footnotesize faces};

\end{tikzpicture}
\caption{A top-level view of the services implemented in \rce.}
\label{fig:rce-architecture}
\end{figure}

As of version 10.0, \rce has more than 230 service interfaces.
We only present an overview of its major functional groups in Figure~\ref{fig:rce-architecture}.
In that figure, dark gray boxes denote the service groups implemented as part of \rce, whereas lightly shaded boxes denote external dependencies.
These external dependencies include the Java runtime itself, but also libraries provided by the Eclipse Rich Client Platform, as well as other third-party-libraries.
In addition to the services shown in Figure~\ref{fig:rce-architecture}, \rce also contains general utilities for authorization, multithreading, logging, testing, installation of updates, and management of multiple user-defined profiles.

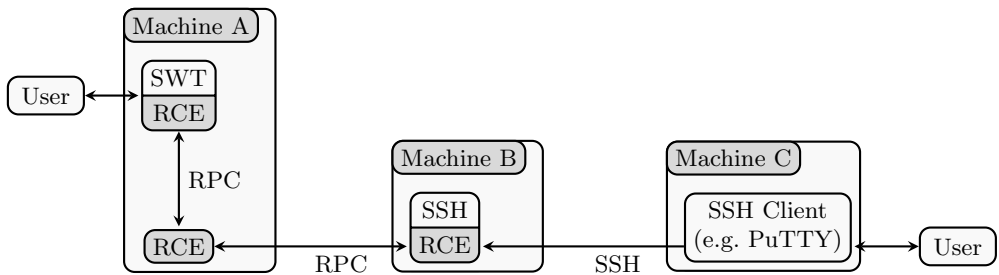
\begin{figure}
\centering
\begin{tikzpicture}

\tikzset{
	user/.style = {
		thick,draw, rounded corners, fill=lightgray,
		minimum width = 1cm,
		minimum height = .5cm,
		path picture = {
			\node at (path picture bounding box.center) {User};
		}
	},
	network connection/.style = {stealth-stealth},
	rpc connection/.style = {stealth-stealth},
	rcebase/.style = {
		thick,
		draw,
		rounded corners,
		fill=red!25,
		minimum width = 1.5cm,
		minimum height= .6cm
	},
	rce/.style = {
		rcebase,
		path picture = {
			\node at (path picture bounding box.center) {\rce};
		}
	},
	rce with interface/.style = {
		rcebase,
		path picture = {
			\node at ($(path picture bounding box.south) + (0,.3cm)$) {\rce};
			\node[draw=blue!50,rounded corners,fill=blue!25,minimum height = .4cm, minimum width=1.2cm] at ($(path picture bounding box.north) - (0,.3cm)$) {#1};
		},
		minimum height= 1.1cm
	},
	machine/.style = {
		thick,
		draw=black,
		fill=black!10,
		rounded corners,
		minimum width = 4cm,
		minimum height = 1cm,
		path picture = {
			\node[minimum width=4cm] at ($(path picture bounding box.north west) + (.9cm, -.3cm)$) {Machine #1};
		}
	}
}

\newcommand{\machine}[2]{
	\begin{pgfonlayer}{background}
		\node[fit={#2},draw,rounded corners,fill=lightgray] (machine) {};
		\node[anchor=south west,draw,rounded corners,fill=darkgray] (machine-label) at (machine.north west) {Machine #1};
		\node[fit={#2 (machine-label)},draw,rounded corners,fill=lightgray,thick] (machine) {};
		\node[anchor=north west,draw,rounded corners,fill=darkgray,thick] (machine-label) at (machine.north west) {Machine #1};
		
	\end{pgfonlayer}
}

	\node[rectangle split, rectangle split parts=2,rectangle split part fill={lightgray,darkgray},draw, text centered,rounded corners,thick] (rce-a) at (0,2) {SWT \nodepart{second} RCE};
	\node[draw,rounded corners,fill=darkgray,thick] (rce-b) at (0,0) {RCE};
	
	\machine{A}{(rce-a) (rce-b)}
	
	\node[user] (user-a) at (-1.75,2) {};
	
	\path[thick] (user-a) edge[<->] (rce-a);
	\path[thick] (rce-a) edge[<->] node[anchor=west] {\footnotesize RPC} (rce-b);
	
	\node[rectangle split, rectangle split parts=2,rectangle split part fill={lightgray,darkgray},draw, text centered,rounded corners,thick] (rce-c) at (3.5,.25) {SSH \nodepart{second} RCE};
	
	\machine{B}{(rce-c)};
	
	\coordinate (rce-c-west) at (rce-c.west);
	\path[<->,draw,thick] (rce-b) -- node[anchor=north,pos=.65] {\footnotesize RPC} (rce-b -| rce-c-west);
	
	\node[draw,rounded corners,thick,align=center,fill=lightgray] (ssh-c) at (7.75, .25) {SSH Client\\(e.g. PuTTY)};
	
	\machine{C}{(ssh-c)};

	\coordinate (ssh-c-west) at (ssh-c.west);
	\coordinate (rce-c-east) at (rce-c.east);
	
	\path[->,draw,thick]
		(ssh-c-west |- rce-b) --
			node[anchor=north,pos=.325] {\footnotesize SSH}
			(rce-c-east |- rce-b);
			
	\coordinate (east-end) at (10.25,.25 |- rce-b);
			
	\node[user] (user-b) at (east-end) {};
	
	\coordinate (user-b-west) at (user-b.west);
	\coordinate (ssh-c-east) at (ssh-c.east);
	
	\path (user-b-west |- rce-b) edge[<->,draw,thick] (ssh-c-east  |- rce-b);
\end{tikzpicture}
\caption{A typical setup of \rce instances and the communication between them.}
\label{fig:rce-communication}
\end{figure}

For communication between different instances, we also use service interfaces to define remote procedure calls (RPCs), which help us in achieving the final requirement above.
On a source code level, a call to a remote procedure is indistinguishable from a call to a local one, thus reducing complexity for software developers.
In Figure~\ref{fig:rce-communication} we present an example of the deployment and usage of \rce as well as the communication between the individual instances.

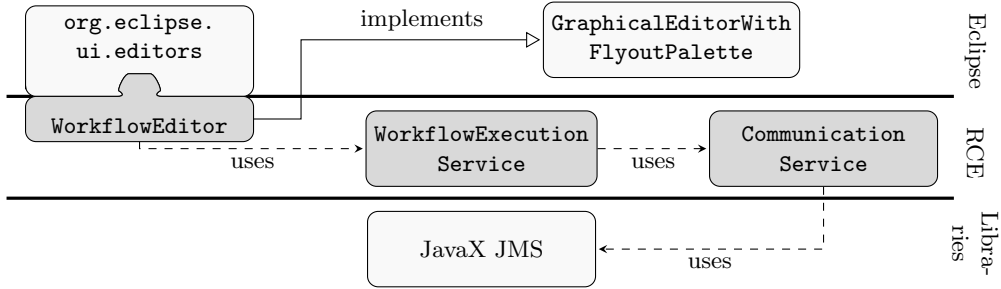
\begin{figure}
\centering
\begin{tikzpicture}[yscale=.6]

\definecolor{servicecolor}{HTML}{83bfe2}
\definecolor{dependencycolor}{HTML}{fde693}

\tikzset{
	box/.style = {
		draw,
		align=center, rounded corners,
		minimum width = 3cm,
		minimum height = 1cm,
		font=\footnotesize
	},
	rce/.style = {
		box,
		draw=black,fill=black!15
	},
	eclipse/.style = {
		box,
		draw=black,fill=gray!5
	},
	implements/.style = {
		draw=black,-{open triangle 60}
	},
	uses/.style = {
		draw=black,dashed,-stealth
	}
}

\draw[very thick] (-1.75,1.75) -- (10.75,1.75);

\draw[very thick] (-1.75,-.5) -- (10.75,-.5);

\node[eclipse] (graphical-editor) at (7,3) {\texttt{GraphicalEditorWith}\\ \texttt{FlyoutPalette}};

\begin{scope}[shift={(-1.5,.75)}]
\begin{scope}[shift={(0,1)}]
\draw[fill=gray!5,rounded corners] (0,0) -- (1.35,0) to[out=135,in=180] (1.5,.5) to[out=0,in=45] (1.65,0) -| (3, 2) -| cycle;
\node[anchor=north,align=center,font=\footnotesize] (extension-point) at (1.5,2) {\texttt{org.eclipse.}\\ \texttt{ui.editors}};
\end{scope}

\draw[fill=black!15,rounded corners] (0,1) -- (1.35,1) to[out=135,in=180] (1.5,1.5) to[out=0,in=45] (1.65,1) -| (3, 0) -- (0,0) -- cycle;
\node[anchor=south,align=center,font=\footnotesize] (workflow-editor) at (1.5,0) {\texttt{WorkflowEditor}};
\end{scope}

\node[rce,anchor=south] (execution-service) at (4.5,-.25) {\texttt{WorkflowExecution}\\ \texttt{Service}};
\node[rce,anchor=south] (communication-service) at (9,-.25) {\texttt{Communication}\\ \texttt{Service}};

\path[implements] (1.5,1.25) -| (2.125,1.75) |- node[near end, anchor=south,font=\footnotesize] {implements}  (graphical-editor);
\path[uses] (0,.75) |- node[near end,anchor=north,align=center,font=\footnotesize] {uses} (execution-service);
\path[uses] (execution-service) -- node[anchor=north,font=\footnotesize] {uses} (communication-service);

\node[eclipse] (jms) at (4.5,-1.625) {JavaX JMS};

\path[uses] (communication-service) |- node[near end,anchor=north,font=\footnotesize] {uses} (jms);

\node[rotate=-90,font=\footnotesize] at (11,2.75) {Eclipse};
\node[rotate=-90,font=\footnotesize] at (11,.5) {RCE};
\node[rotate=-90,font=\footnotesize,align=center] at (11,-1.5) {Libra-\\ ries};

\end{tikzpicture}
\caption{Excerpt of the services involved in the execution of a workflow via the GUI.}
\label{fig:architecture-workflow-execution}
\end{figure}

Due to space limitations we are unable to describe all service interfaces and their dependencies in detail.
Instead, we conclude this section with an overview over the implementation of a typical use case, namely over the construction and execution of a workflow via the GUI.
We illustrate the services involved and the communcation between them in Figure~\ref{fig:architecture-workflow-execution}.

In order to create and modify workflows, \rce provides users with a graphical editor.
This editor is implemented in the class \texttt{Work\-flow\-E\-di\-tor}, which extends the abstract class \texttt{Gra\-phi\-cal\-E\-di\-tor\-With\-Fly\-out\-Pa\-lette}.
The latter class is part of the Eclipse RCP framework.
We furthermore register \rce's \texttt{WorkflowEditor} with the Eclipse framework using the extension point \texttt{org.eclipse.ui.editors}.
This registration instructs Eclipse to instantiate a \texttt{WorkflowEditor} if the user wants to edit a workflow file, which are identified via their file ending \texttt{.wf}.

Once the user has finished construction of her workflow, she executes it using a button in the workflow editor.
\rce passes the constructed workflow to the \texttt{WorkflowExecutionService}.
That service initializes all instances involved in the execution of the workflow, starts all tools that do not require inputs, collects the output data of tools, determines the subsequent tools to execute, and passes the data to the inputs of those tools.
In order to communicate with remote instances of \rce, the \texttt{WorkflowExecutionService} uses the \texttt{Com\-mu\-ni\-ca\-tion\-Ser\-vice}.
That service encapsulates JavaX JMS, which it uses for the actual communication between instances of \rce.

When a service is instantiated, e.g., because the user opens a workflow file using the GUI and the Eclipse framework instantiates a \texttt{WorkflowEditor}, OSGi is used to inject the dependencies of the \texttt{WorkflowEditor} into the constructed instance.
After having given an overview over both the functionality of the software as well as its internal architecture, we now continue with an example of how \rce might be used in a typical engineering project.

\section{Concrete Example}
\label{sec:illustrative_examples}

\begin{figure}
\centering
\input{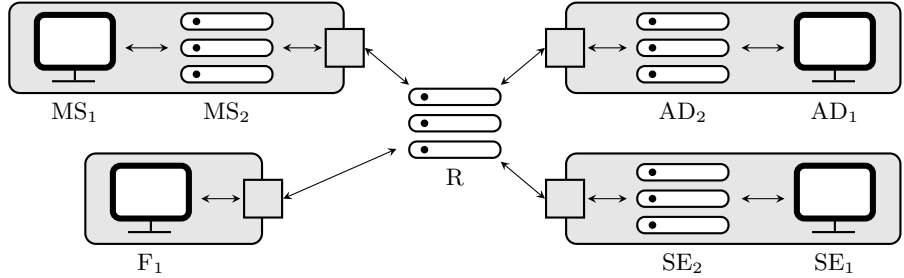}
\caption{A network setup of the machines taking part in the process shown in Figure~\ref{fig:initial-workflow}.}
\label{fig:network-setup}
\end{figure}

Consider again the example of the design of an airplane wing from Section~\ref{sec:motivation_and_significance} and recall that the individual tools implementing the steps of the process as illustrated  in Figure~\ref{fig:initial-workflow} may be required to run on different machines.
Although these machines may be part of different disjoint networks, we assume that the networks themselves are connected via the Internet.
Otherwise, no communication is possible.
We illustrate such a network setup in Figure~\ref{fig:network-setup}.

In that drawing, the vertices labeled \MS, \AD, \F, and \SE denote machines under the control of the department for materials science, aerodynamics, finance, and structural engineering, respectively.
Furthermore, the vertex labeled \R represents a dedicated relay server in a network location that is reachable from all project participants.
We denote these isolated networks by dark gray boxes in Figure~\ref{fig:network-setup}, each of which has an interface connecting it to the relay server.

Each involved project partner starts by defining the integration of their respective tool, as described in Section~\ref{sec:software_description:functionalities}, on their local computer (i.e., on the machines $\MS_1$, $\AD_1$, $\SE_1$, and $\F_1$).
This may happen either using the graphical wizard for tool integration, or via manually editing the files defining the tool integration.
After completing and testing the integration locally, the users move the integration to the compute nodes (i.e., to the machines labelled $\MS_2$, $\AD_2$, and $\SE_2$).
For simpler tools, in contrast, the users may opt to leave both the tool as well as the integration on their local computers.
In our example, this is the case for the finance department, which does not use a compute node.

Each machine that hosts one or more tools is connected to the central relay server~$\R$, thus making them visible to one another.
The users then collaborate in constructing the workflow shown in Figure~\ref{fig:rce-workflow}, which formalizes the workflow illustrated in Figure~\ref{fig:initial-workflow} on Page~\pageref{fig:initial-workflow}.
Within this workflow, they can transparently mix tool instances that were moved to a compute node with tools that are still published from a user's machine.

Having finished the construction of the workflow, the users may execute it using, e.g., the relay server or one of the compute nodes as the controller for the distributed execution.
Using such a shared server as the workflow controller allows machines not included in the actual computation (i.e., the machines~$\MS_1$,~$\AD_1$, and~$\SE_1$) to be safely turned off or disconnected from the \rce network during the execution of the workflow.
This is particularly useful for machines that are not permanently attached to the same physical network, e.g., work laptops switching between cable and wireless adapters.

\section{Impact}
\label{sec:impact}

\rce is widely used among many projects at DLR involving users from the fields of aeronautics and astronautics, as well as traffic and energy.
The specific impact of the use of \rce on the performance of an engineering project is hard to measure due to multiple and conflicting definitions of this notion.
However, numerous users have lauded particular features of \rce that have of great use in their projects, such as the capability to create repeatable and reusable workflows, or the ability to seamlessly perform simulations on a distributed network~\cite{BodenFlinkMischkeEtAl2019B}.
In addition to this reported increase in efficiency, multiple users rely on the use of \rce in their daily work and report of successfully executed engineering projects which would not have been possible without the use of \rce in the first place.

Since a complete overview over all projects that involve \rce is impossible due to its nature as freely distributed open-source software, we highlight a number of representative projects here.

One of the major application domains of \rce is the domain of aerospace engineering.
Boden, Flink, Mischke, et al.~\cite{BodenFlinkMischkeEtAl2019B} give an overview over such projects.
Here, we highlight two such projects, namely \freacs and \digitalx, as well as a number of projects in other domains that leveraged the capabilities of \rce.

\subsection{\freacs}

The goal of the project~\freacs~\cite{MoerlandPfeifferBoehnkeEtAl2017} was to design a product development process for aircraft configurations.
To this end, the engineers involved defined engineering services, i.e., disciplinary tools that were required by other participants in the project.
Both the publication as well as the connection of the published engineering services were done via \rce.
Since, in this development setting, each engineering service was owned and maintained by a specialist engineer, the project benefited greatly from the use of \rce, as each service and its executions remained under complete control of their owners.
The resulting data was organized using the integrated project explorer of \rce, which allowed for quick analysis of the data generated by running the individual services.

A major part of the project \freacs consisted of regularly occurring design camps, in which all project participants gathered in a single location to further design and enhance the development process.
These design camps not only greatly benefited from \rce's ability to export the accrued data already during the execution of the workflow, but also from the easy access to the data from all instances of \rce that are connected to the network.
This allowed the users to detect and fix errors early in the design phase of the workflow.

\subsection{\digitalx}

\rce was also used in the DLR project \digitalx~\cite{GoertzIlicAbuZuraykEtAl2016,GoertzIlicJepsen2017}.
While previously, \rce workflows were mainly used in the low-fidelity pre-design of aircraft, the aim of this project was the development of a multidisciplinary optimization process that comprised both low- and high-fidelity simulations of aircraft.
To this end, a number of highly specialized tools running on either Windows or Linux were integrated into a single workflow.
These tools were contributed by eight DLR institutes and hosted at six different locations.
Hence, the use of \rce in this project showcases the ease of integrating tools requiring different runtime environments into a single workflow.

Moreover, although starting the high-fidelity analysis required the results of the low-fidelity analysis, the individual steps of the former and the latter analyses were partly independent of previous results.
Thus, the performance of these individual steps improved markedly over previous implementations of this workflow, as \rce automatically executes independent tools in parallel~\cite{GoertzIlicJepsen2017}.

\subsection{\pegasus}

Apart from its use in the design of complete aircraft, \rce was also employed for the design of individual airplane engines as part of the project \pegasus~\cite{ReitenbachKrummeBehrendtEtAl2018}.
In this project, a special integration interface was developed that allowed an existing \cpp software to connect to \rce instances using the standardized SSH protocol.
Leveraging \rce's existing network capabilities, this allowed that software to execute distributed simulation tools, and also invoke other instances of the \cpp software itself.
The latter was also done to improve performance by distributing compute loads across multiple machines, which required system information to do load balancing.
To support this, \rce made the performance data that it already collects available over the SSH interface.
To simplify usage of these features, all communication with \rce was encapsulated in a client library providing a C API, which allows it to be used from a wide range of programming languages.
This library is currently being modernized and is planned for inclusion in future \rce releases.

\subsection{\triad}

In addition to its use in numerous aerospace projects, \rce was furthermore used in the project \triad~\cite{WeiandBuchwaldSchwinn2019} which focused on developing a design environment for helicopters.
Here, again multiple individual disciplinary tools were integrated into a single workflow comprising both low- and high-fidelity analysis and optimization of rotorcraft.
Since the aim of the project was not to develop a single rotorcraft, but rather to provide a general environment with which multiple such craft could be designed, there was no fixed workflow to be implemented.
Instead, the participating engineers integrated their tools into \rce and published their tools onto the network.

One of the major work items of this project was the development of the actual workflow.
As its structure was not clear at the beginning of the project, \rce's intuitive graphical interface significantly reduced the workload of the individual engineers.
Moreover, several of the involved tools were proprietary and were restricted to running on the machines of certain developers.
Hence, \rce's capability of executing tools on the publishing machine itself proved to be a great benefit to the construction of the overall workflow.

\subsection{\clava}

Finally, while a large number of known users works in aerospace engineering, \rce is also deployed in the field of astronautics.
Prime among these applications is the use of \rce in the project \clava~\cite{FischerDeshmukhKochEtAl2018}, which has the task of developing the next generation of launchers.
This design process is highly dynamic in nature, since the used tools are regularly replaced by newer versions or other tools entirely.
Hence, the involved engineers were able to speed up the design process via the use of \rce, which transparently uses new versions of the tools as soon as they are deployed and allows for easy replacement of the used tools.
Furthermore, another increase in velocity resulted from the automatic execution of the workflow which was previously executed manually, i.e., via manual execution of the involved tools and dissemination of their outputs.

Due to space restrictions, we omit detailed descriptions of other projects in which \rce is deployed or currently involved.
These projects involve ones from the domains of logistics and climate research (e.g., \trak or \veuII~\cite{SeumEhrenbergerKuhnimhofEtAl2019}), energy research (e.g., \inteeverII), or shipbuilding (e.g., \holiship~\cite{Papanikolaou2019}).
Moreover, since \rce is freely available as open source software, we do not have access to a comprehensive list of users.

\section{Conclusion and Future Work}
\label{sec:conclusion}

In this work we have presented \rce (Remote Component Environment), an open-source integration environment for engineers and scientists that aids them in the multidisciplinary development, analysis, and the optimization of complex systems.
We have given an overview over the main features of \rce and argued that due to the richness of this feature set, \rce is able to adapt to numerous usage scenarios.
Instead of having to adapt to a fixed design process prescribed by \rce, engineering teams can use \rce to enhance their existing design processes.
Subsequently, we have outlined the technical implementation of these features and we have shown that our choice of frameworks, development methodology, and supporting technologies enables not only the further maintenance of \rce, but also the development of additional features as required by future projects.

Finally, we have given an overview over research and development projects at \dlr that use or have used \rce as part of their engineering processes.
These projects showcase the versatility of \rce, as it is not only used in a number of domains, but since it also supports a wide range of scales and scopes of projects.
This includes both low- and high-fidelity analyses, simulations, and optimizations, as well as different modes of interaction, ranging from iterative design work using the graphical interface up to completely automated invocations of \rce via its network interfaces.
In all these projects, users report increased productivity through using \rce and that in some cases \rce has enabled their project in the first place.

For future work, we will further improve the usability of \rce and increase its feature set based on requirements arising from the use in scientific and engineering projects.
Prime among these future developments is an extension and modularization of the central code governing the execution of workflows, which forms one of the oldest and most monolithic parts of the current codebase.
Further, we are adapting \rce for use in cloud environments to allow users to optimally leverage shared resources.
Moreover, as there is an increasing need for sharing integrated tools between cooperating organizations, dedicated support for transfer of sensitive data over insecure communication channels, such as the Internet, will be added in the near future.
To address the IT security demands of such setups, the network protocol for this will be encrypted, authenticated, and restricted to a minimal and easily audited feature set.
Finally, due to the ever-increasing size of workflows, one of our main goals is to provide users with improved editing and management features for larger workflows.
This includes both the capability to modularize parts of workflows into sub-workflows, as well as the ability to publish complete workflows as virtual tools, which can then be used in other workflows just like standard tools.
Together, these features will improve the design as well as simplify testing and maintenance of large-scale workflows.

\paragraph*{Acknowledgements}
\label{sec:conclusion:acknowledgements}

We gratefully acknowledge the support and contributions of previous team members Doreen Seider, Arne Bachmann, Tobias Brieden, Markus Kunde, Markus Litz, Oliver Seebach, Heinrich Wendel, and Sascha Zur.
Furthermore, we would like to thank all other colleagues, students, and users that have  contributed to the development of \rce.

\bibliographystyle{splncs04} 
\bibliography{main}

\end{document}